\documentclass[12pt]{report}
\usepackage[dvips]{graphicx}
\newcommand{\sptwo}{1.4}

\newcommand{\doublespace}{\edef\baselinestretch{\sptwo}\Large\normalsize}

\begin{document}
\doublespace

\begin{center}
{\bf Cold Bose Gases near Feshbach Resonances}
\\
\renewcommand\thefootnote{\fnsymbol{footnote}}
{Yeong E. Kim \footnote{ e-mail:yekim$@$physics.purdue.edu} and
Alexander L. Zubarev\footnote{ e-mail: zubareva$@$physics.purdue.edu}}\\
Purdue Nuclear and Many-Body Theory Group (PNMBTG)\\
Department of Physics, Purdue University\\
West Lafayette, Indiana  47907\\
\end{center}
\begin{quote}
The lowest order constrained variational method [Phys. Rev. Lett. {\bf 88},
210403 (2002)] has been generalized for a dilute
 (in the sense that the
range of interatomic potential, $R$,  is small compared with inter-particle
spacing $r_0$) uniform gas of bosons near the Feshbach resonance using the
 multi-channel zero-range potential model. The method has been applied to $Na
 (F=1, m_F=1)$ atoms near the $B_0=907$G Feshbach resonance. It is shown that
 at high densities $n a^3\gg 1$, there are  significant differences between our results
 for the real part of energy per particle and the one-channel zero-range
 potential
 approximation. We point out the possibility of stabilization of the uniform condensate for the case of negative scattering length.
\end{quote}

\vspace{5mm}

\vspace{55 mm}
\noindent

\pagebreak

{\bf I. INTRODUCTION}
\vspace{8pt}

The newly created Bose-Einstein condensates (BEC) of weakly interacting
alkali-metal atoms [1] stimulated a large number of theoretical investigations
(see recent reviews [2]).
Most of these works are based on the assumption that  the  properties of BEC
are well described by the Gross-Pitaevskii (GP) mean-field theory [3].

Recently, it has become possible to tune atomic scattering length to
 essentially any value, by exploiting Feshbach resonances (FR) [4,5].
 A fundamental open
problem is how to describe the physics of dilute BEC near FR (dilute in the
 sense that the 
range of interatomic potential, $R$,  is small compared with inter-particle 
spacing $r_0$)  in the regime of
a large scattering length, $a$, which we take to be positive.
The GP approach fails in the regime of a large gas parameter, $n\mid a \mid^3$, where $n$ is the particle density.

The dilute BEC for a large gas parameter regime in one-channel approximation 
has been considered previously in Ref.[6] (for the corresponding problem for 
a Fermi gas see Ref.[7]).

In this paper we consider the ground state properties of the dilute
 homogeneous Bose gas near the FR using a multichannel zero-range potential
 (ZRP)
 model of FR.

In  section II we describe the lowest order constrained variational (LOCV)
 method [6,8] for the one-channel N-body problem. The calculations for model
 interaction potentials used by Ref.[9] are presented. The description of the
 multi-channel ZRP model is given in  section III.
 Section IV develops the LOCV method for the dilute Bose gas near FR. 
We conclude the paper in  section V with a brief summary.
 
\vspace{8pt}

{\bf II. LOWEST ORDER CONSTRAINED VARIATIONAL METHOD}
\vspace{8pt}

In a dilute many-body problem in the large gas parameter regime correlations
 between particles are very important.

The LOCV method [6,8] for the homogeneous N-body system is to assume a Jastrow
 many-body wave function, of the
 form
$$
\Psi(\vec{r}_1, \vec{r}_2,... \vec{r}_N)=\prod_{i<j} f(\vec{r}_i-\vec{r}_j),
\eqno{(1)}
$$
where at short distances, $f$ is solution of the two-particle Schr\"odinger
equation 
$$
(-\frac{\hbar^2}{m}\frac{d^2}{d r^2}+V(r) )r f(r))=\lambda r f(r),
\eqno{(2)}
$$
while at large distances $f$ must approach a constant.

In the LOCV method [6,8] the boundary conditions for $f(r)$ are
$$f(d)=1,\: f^{\prime}(d)=0,
\eqno{(3})
$$
the expectation value of the energy is given by
$$
E/N=2 \pi n \lambda \int_0^d f^2(r) r^2 dr,
\eqno{(4)}
$$
and $d$ is defined by the normalization
$$
4 \pi n \int_0^d f^2(r) r^2 dr=1.
\eqno{(5)}
$$

In Ref.[6] for the dilute case ($R\ll r_0$, where the inter-particle spacing 
$r_0=(3/(4 \pi n))^{1/3}$) inter-atomic interaction was replaced by the zero-range potential (ZRP) model [9]
$$
\frac{(rf)^{\prime}}{rf}\big{|}_{r=0}=-\frac{1}{a}.
\eqno{(6)}
$$
In this case 
$$
f(r<d)=\frac{d}{r} \frac{\sin(kr+\delta)}{\sin(kd+\delta)},
\eqno{(7)}
$$
with $k \cot\delta=-1/a$, $~$ $kd \cot(kd+\delta)=1$,
where $k=\sqrt{m \lambda/\hbar^2}$.

In the small $a/r_0$ limit, $~$ $\delta=-ka$, $~$ $d=r_0$ and the LOCV result for $E/N$ [6] is given by
$$
\frac{E}{N}=2 \pi \frac{\hbar^2 a}{m} n
\eqno{(8)}
$$
and is same as that first found by Lenz [10].

The ground state energy per particle, $E/N$, in the low-density regime,
$n a^3 \ll 1$,
can be calculated using an  expansion in power of $\sqrt{n a^3}$
$$ 
\frac{E}{N}=\frac{2 \pi \hbar^2}{m} an[1+\frac{128}{15 \sqrt{\pi}}
(n a^3)^{1/2}+8 (\frac{4 \pi}{3}-\sqrt{3}) na^3 [\ln(n a^3)+C]+...].
\eqno{(9)}
$$
The coefficient of the $(na^3)^{3/2}$ term (the second term) was first 
calculated by Lee, Huang, and Yang [11], while the coefficient of the last term 
was first obtained by Wu [12]. The constant $C$ after the logarithm was 
considered
 in Ref.[13].

The expansion (9) is asymptotic, and it was shown in Ref.[14] that the
Lee-Huang-Yang (LHY) correction (second term in Eq.(9)) represents a
significant improvement on the mean field prediction (the first term in Eq.(9))
.

\pagebreak

In Refs.[15-19] the Lenz-Lee-Huang-Yang (LLHY) expansion (first two terms
in expansion (9)) has been used to study effects beyond the mean field
 approximation. In Ref.[18], it was found that the correction to the GP results
 may be as large as 30\% in the ground state properties of the condensate,
 when the conditions of the JILA experiment for $^{85}Rb$ are considered [5].

In the large $a/r_0$ limit, $\delta=\pi/2$, $kd \tan (kd)=-1$, and the energy is
 given by [6]
$$
\frac{E}{N}=13.33 \hbar^2 \frac{n^{2/3}}{m}=10.2597 \frac{\hbar^2}{2 m r_0^2},
\eqno{(10)}
$$
which is very close to the Legett's unpublished variational result
$$
\frac{E}{N}=\frac{3 \pi^2}{2} \frac{\hbar^2 n^{2/3}}{m}
\eqno{(11)}
$$
(this was quoted by Baym [20]).

To study the validity of the ZRP model we consider an example of the
 square-well (SW)  potential
with $a/r_0 \rightarrow \infty$. The calculated energies per
 particle, $E/N$, are presented in Table I. The ratio $R/r_0$ is typically of
 the order of $10^{-2}$. From Table I, we can see that even for 
$a/r_0 \rightarrow \infty$ the ZRP model is a very good approximation
 (the difference for $E/N$ between the ZPR and the SW is less than 1\% if
 $R/r_0\approx 10^{-2}$). In this case the LOCV results for  $E/N$ have
 universal properties that depend on the interatomic potential only through the single low-energy parameter $a$, even for large gas-parameter regime.

We note that for an attractive potential the atomic BEC is metastable and
energy per particle can be written as $E/N-i\Gamma/2$. Therefore the 
real
 part of energy in
LOCV method 
  would be reliable if $E/N>\Gamma/2$ [6]. 

In Ref.[14], the authors have used a diffusion Monte-Carlo method to calculate 
the lowest-energy state of uniform gas of bosons interacting through four 
different model potentials  that have all the same scattering length $a$.

 (i) The hard-sphere potential
$$
V^{(HS)}(r)=\left\{\begin{array}{ll}
\infty&\mbox{$(r<a$)}\\
0&\mbox{$(r>a)$,}
\end{array}
\right.
\eqno{(12})
$$
where the diameter of the hard-sphere is equal to the scattering length.

(ii) Two soft-sphere (SS) potentials
$$
V^{(SS)}(r)=\left\{\begin{array}{ll}
V_0&\mbox{$(r<R_0)$}\\
0&\mbox{$(r>R_0)$,}
\end{array}
\right.
\eqno{(13})
$$
with $R_0=5 a$, $~$ $V_0=0.031543 \hbar^2/(m a^2)$ $~$ and
$R_0=10 a$, $~$ $V_0=0.003408 \hbar^2/(m a^2)$.

(iii) The hard-core square-well (HCSW) potential
$$
V^{(HCSW)}(r)=\left\{\begin{array}{lll}
+\infty&\mbox{$(r<R_c)$}\\
-V_0 &\mbox{$(R_c<r<R_0)$}\\
0&\mbox{$(r>R_0)$.}
\end{array}
\right.
\eqno{(14})
$$
The parameters of the HCSW potential are $R_c=a/50$, $R_0=a/10$ and $V_0=
412.815 \hbar^2/(m a^2)$. The HCSW potential has a two-body bound state 
with energy $-1.13249 \hbar^2/(m a^2)$ [21].

Comparison the LOCV results for potentials (12-14) with the available 
diffusion Monte-Carlo (DMC) calculations (Table II, Fig.1 and Fig.2) shows that the
 LOCV energies in the case of $R/r_0 \approx 10^{-2}$ are in very good
agreement  with the DMC results.



\vspace{8pt}

{\bf III. ZERO-RANGE POTENTIAL MODEL OF FESHBACH RESONANCE}
\vspace{8pt}

We  start from the coupling channel equation
$$
\everymath={\displaystyle}
\begin{array}{rcl}
(-\frac{\hbar^2}{m} \nabla^2+V^P(r))\psi^P(\vec{r})+g(r) \psi^Q(\vec{r})=E
 \psi^P(\vec{r}),\\ \\
(-\frac{\hbar^2}{m} \nabla^2+V^Q(r)+\mathcal{E})\psi^Q(\vec{r})+g^{\ast}(r)
 \psi^P(\vec{r})=E
 \psi^Q(\vec{r}),
\end{array}
\eqno{(15)}
$$
where $\mathcal{E}$ is the energy shift of the closed channel $Q$ with respect
 to the collision continuum.
 Since potentials $V^P(r)$, $V^Q(r)$ and $g(r)$ have  a range typically of
 the order of
 interatomic potential range  or less,  we can replace
 Eq.(15) by
$$
\everymath={\displaystyle}
\begin{array}{rcl}
(-\frac{\hbar^2}{m} \nabla^2+V_{11})\psi^P(\vec{r})+ V_{12}\psi^Q(\vec{r})=E
 \psi^P(\vec{r}),\\ \\
(-\frac{\hbar^2}{m} \nabla^2+V_{22}+\mathcal{E})\psi^Q(\vec{r})+V_{21}
 \psi^P(\vec{r})=E
 \psi^Q(\vec{r}),
\end{array}
\eqno{(16)}
$$
where
$$
V_{ik}(\vec{r})=-\frac{4 \pi \hbar^2}{m} M^{-1}_{ik} \delta(\vec{r})
 \frac{\partial}{\partial r} r
\eqno{(17)}
$$
is  a multichannel generalization of the Huang pseudo-potential [22].

Equations (16) can be rewritten as free equations
$$
\everymath={\displaystyle}
\begin{array}{rcl}
-\frac{\hbar^2}{m} \nabla^2 \psi^P(\vec{r})=E
 \psi^P(\vec{r}),\\ \\
(-\frac{\hbar^2}{m} \nabla^2+\mathcal{E})\psi^Q(\vec{r})=E
 \psi^Q(\vec{r}),
\end{array}
\eqno{(18)}
$$ 
with boundary conditions
$$
\everymath={\displaystyle}
\begin{array}{rcl}
\frac{d(r \psi^P)}{d r} |_{r=0}=(M_{11}r\psi^P+M_{12}r\psi^Q)|_{r=0},\\ \\
\frac{d(r \psi^Q)}{d r} |_{r=0}=(M_{21}r \psi^P+M_{22}r\psi^Q)|_{r=0},
\end{array}
\eqno{(19)}
$$
which is a multi-channel ZRP model  [23].

It is easy to show that the Hamiltonian of the particle moving in the field of multi-channel ZPR is Hermitian, that is, the condition
$$
\int[(\phi_1^P)^{\ast}\nabla^2\phi_2^P+(\phi_1^Q)^{\ast}\nabla^2\phi_2^Q]d^3 r
=\int[(\nabla^2 \phi_1^P)^{\ast} \phi_2^P+(\nabla^2 \phi_1^Q)^{\ast} \phi_2^Q]
d^3 r
\eqno{(20)}
 $$
is valid for any $\phi_1^P, \phi_1^Q$ and $\phi_2^P, \phi_2^Q$ which satisfy the boundary conditions (19) with energy independent  constant Hermitian
 matrix $M$.

We shall use  the following parameterization
 for the matrix $M$ 
$$
M_{11}=-\frac{1}{a_{bg}},\: M_{12}=M_{21}=-\frac{\beta}{a_{bg}}, \:
M_{22}=-\frac{\gamma}{a_{bg}},
\eqno{(21)}
$$
where $a_{bg}$ is the background value of the scattering length. The model
(21) with $\gamma=1$ was considered in [23].

Solutions of Eqs.(18,19) can be written as
$$
\everymath={\displaystyle}
\begin{array}{rcl}
\psi^P(\vec{r})\propto \sin(k r+\delta),\\ \\
\psi^Q(\vec{r})\propto e^{-\sqrt{\tilde{\mathcal{E}}-k^2}r},
\end{array}
\eqno{(22)}
$$
with $\tilde{\mathcal{E}}=m \mathcal{E}/\hbar^2$, $k^2=m E/\hbar^2$.

Since the energy shift $\mathcal{E}$ can be converted into an external magnetic field
 $B$ by $\mathcal{E} \propto B$, the scattering length $a$ depends
 on the external magnetic field by the dispersive law [24]
$$
a=a_{bg} (1+\frac{\Delta(B)}{B_0-B}),
\eqno{(23)}
$$
where
$$
\Delta(B)=\frac{(1-\sqrt{\alpha B_0})(\sqrt{\alpha B}+\sqrt{\alpha B_0})}
{\alpha},
\eqno{(24)}
$$
$$
\alpha=\frac{4 B_0}{(2 B_0+\Delta)^2}, \:
\frac{\beta}{\gamma}=\sqrt{\frac{\Delta}{\Delta+2 B_0}},
\eqno{(25)}
$$
and $\Delta=\Delta(B_0)$ characterizes  the resonance width.

Two parameters of our model Eq.(21), $a_{bg}$ and $\beta/\gamma$, are completely
 determined from the external magnetic field dependence of the scattering
 length. For example, in the sodium case [25] the width $\Delta$ of the   
Feshbach resonance ($B_0=907$G) is $1$G, therefore
 $\alpha=1.10132 10^{-3}1/$G, $\beta/\gamma=2.34726 10^{-2}$, and $a_{bg}=53$au.

As for parameter $\gamma$, it can be determined from the energy dependence of 
the scattering phase shift, $\delta$ 
$$
k \cot \delta=\frac{1}{a_{bg}} (-1+\frac{\Delta}
{\Delta+2 B_0-\sqrt{4 B_0 B-(k a_{bg} (\Delta+2 B_0)/\gamma)^2}}).
\eqno{(26)}
$$

We note that the multi-channel ZRP model is different from the contact 
potential model
of Ref.[26], which uses a cutoff regularization in the momentum representation.

\vspace{8pt}

{\bf IV. DILUTE BOSE GAS NEAR FESHBACH RESONANCE}
\vspace{8pt}

For a  generalization of the LOCV method for a  dilute uniform gas of bosons 
interacting through the two-channel ZRP, Eqs.(18,19,21), we assume
$$
\Psi^P(\vec{r}_1, \vec{r}_2,... \vec{r}_N)=\prod_{i<j} f^P(\vec{r}_i-\vec{r}_j),
\eqno{(27)}
$$
where $\Psi^P(\vec{r}_1, \vec{r}_2,... \vec{r}_N)$ is the Jastrow wave function,  index $P$ denotes the projection onto the Hilbert subspace of the incident
(atomic) channel, and $f^P$ at short distance is solution of the Schr\"odinger
equation
$$
-\frac{\hbar^2}{m} \frac{d^2}{d r^2}rf^P(r)=\lambda rf^P(r),
\eqno{(28)}
$$
with boundary conditions
$$
\everymath={\displaystyle}
\begin{array}{rcl}
\frac{d(r f^P)}{d r} |_{r=0}=-\frac{1}{a_{eff}}rf^P,\\ \\
f^P(d)=1,\: \frac{df^P}{d r}|_{r=d}=0,
\end{array}
\eqno{(29)}
$$
where
$$
a_{eff}=a_{bg}(1+\frac{\Delta}{2B_0-\sqrt{4 B_0 B-(\kappa a_{bg} (\Delta+2 B_0)/
\gamma)^2}}),
\eqno{(30)}
$$
and $\kappa=\sqrt{m \lambda}/\hbar$. 
The real part of ground-state energy per particle is given by
$$
E/N=2 \pi n \lambda \int_0^d (f^P(r))^2 r^2 dr,
\eqno{(31)}
$$
 and $d$ is defined by the normalization
$$
4 \pi n \int_0^d (f^P(r))^2 r^2 dr=1.
\eqno{(32)}
$$
 We note that the effective scattering length,
 $a_{eff}$ is a many-body parameter (it depends on $E/N$),
and, for $B=B_0$, $a_{eff}$ does not tend to infinity.

The calculated energies per particle, $E/N$, of $ Na (F=1, m_F=1)$ atoms at
 resonance magnetic field ($B=B_0=907$G) for $r_0=10^2 a_{bg}$ are compared
 with the one-channel
 approximation, Eq.(10), in Fig.3. These comparisons show that for finite
 values of $\gamma$ there are significant differences between our results and
 the approximation of
 Ref.[6]. Our results are much smaller than the Legett's variational estimate
[20].

To consider, so called, nonuniversal effects [21], i.e. the sensitivity to the
 parameters of the interatomic interactions other than the scattering length,
 we calculate $E/N$ as a function of $\gamma$. Table III shows a strong $\gamma$
 dependence of $E/N$ near the FR. 

 However near the FR the atomic BEC is metastable, and the LOCV $E/N$ would be 
reliable if $E/N>\Gamma/2$ [6]. We have extracted the values of $\Gamma/2$ from 
experimental data of Ref.[25]. Using these values of $\Gamma/2$ we have 
calculated  the ratio $(\Gamma/2)/(E/N)$ (see Table IV). From  Table IV we can 
see that the LOCV results for the real part energy, $E/N$, is valid
 for the experimental conditions of Ref.[25], since
 $(\Gamma/2)/(E/N)\approx 10^{-2}$.

Now suppose  that $a$, Eq.(23), near the FR is negative ($a_{bg}$ for the
$ Na (F=1, m_F=1)$ atoms is positive). In one-channel case the uniform 
condensate for negative $a$ is always mechanically unstable. But the two-channel
 consideration can lead to stable uniform solution, since
the many-body parameter  $a_{eff}$ can be
 positive. Table V illustrates this stabilization effect. Although the
 three-body recombination processes [23,25,27-29] can make it difficult to
 observe this effect
 experimentally, we note that Ref.[30]  considered the possibility of
 suppressing three-body recombinations in a trap.
There is a similar case [30]  of uniform 1D gas of N bosons on a ring 
for which
inelastic decay processes, such as the three-body recombination, are
suppressed in the strongly interacting and intermediate regimes.
\vspace{8pt}

\pagebreak

{\bf V. SUMMARY AND CONCLUSION}
\vspace{8pt}

In summary we have considered the LOCV  method for a  dilute uniform gas of
 bosons.

Comparison of the LOCV results for potentials Eqs.(12-14) with the available
diffusion Monte-Carlo (DMC) calculations  shows that
the
 LOCV energies in the case of $R/r_0 \approx 10^{-2}$ are in very good
agreement  with the DMC results.

 We have generalized the LOCV method [6,8] for dilute uniform gas of bosons near the FR, using the multi-channel ZRP model.

As an example of application, we have considered $Na (F=1, m_F=1)$ atoms near
 the $B_0=907$G FR.

At high density $n a^3 \gg 1$, there are  significant
 differences between our results
 and one-channel ZRP [6] 
for the real part of energy per particle.

For the case of negative scattering length, we point out the possibility of
 stabilization of the uniform condensate. 
\pagebreak

TABLE I. Energy per particle, $E/N$, in units of $\hbar^2/(2 m r_0^2)$
vs $R/r_0$ for the square-well attractive potential with $a=\infty$.
The ZRP model for this case gives $E/N=10.2597 \hbar^2/(2 m r_0^2)$ [6].
\vspace{10mm}

\begin{tabular}{ll}
\hline\hline
$R/r_0$
&E/N
 \\ \hline
$10^{-1}$
&11.1906
 \\ \hline
$5 \times 10^{-2}$
&10.7143
 \\ \hline
$10^{-2}$
&10.3502
 \\ \hline
$5 \times 10^{-3}$
&10.3057
 \\ \hline
$10^{-3}$
&10.2703
 \\ \hline\hline
\end{tabular}\\

\vspace{10mm}

TABLE II. Energy per particle, $E/N$, in units of $2 \pi \hbar^2 n a/m$, 
as a function of
 $R/r_0$ for soft-sphere potentials. $R$ is the range of potential and 
$
r_0=(3/(4 \pi n))^{1/3}$ is the atomic separation.
\vspace{10mm}

\begin{tabular}{llll} 
\hline\hline
$R/r_0$
&E/N
&E/N[14]
&R
 \\ \hline
0.081
&1.01961
&1.00427
&$5a$
 \\ \hline
0.161
&1.01951
&1.00427
&$10 a$
 \\ \hline
0.174
&1.04292
&1.01382
&$5 a$
 \\ \hline
0.347
&1.04288
&1.01302
&$10 a$
 \\ \hline
0.374
&1.09599
&1.04167
&$5 a$
 \\ \hline
0.748
&1.0973
&1.03689
&$10 a$
 \\ \hline
0.806
&1.23315
&1.11011
&$5 a$
 \\ \hline\hline
\end{tabular}\\

\vspace{10mm}

\pagebreak

TABLE III. Energy per particle of $Na (F=1, m_F=1)$ atoms, $E/N$, multiplied by $10^6 2 m a_{bg}^2/\hbar^2$
as a function of
 $\gamma$ and $B$ near the $B=907$G Feshbach resonance.
\vspace{10mm}

\begin{tabular}{lllllll}
\hline\hline
$\gamma$
&$B=906$G
&$B=906.5$G
&$B=906.8$G
&$B=906.9$G
&$B=906.98$G
&$B=908.5$G
 \\ \hline
0.1
&5.147
&6.008
&6.835
&7.185
&7.498
&0.889
 \\ \hline
0.5
&6.078
&8.939
&15.59
&21.51
&29.58
&0.998
 \\ \hline
1.0
&6.128
&9.225
&18.06
&29.69
&55.30
&1.002
 \\ \hline
2.0
&6.141
&9.304
&19.00
&34.94
&100.3
&1.003
 \\ \hline
3.0
&6.144
&9.319
&19.20
&36.41
&137.9
&1.003
 \\ \hline
4.0
&6.144
&9.319
&19.27
&36.41
&168.2
&1.003
 \\ \hline
5.0
&6.145
&9.327
&19.30
&37.27
&197.5
&1.003
 \\ \hline
6.0
&6.145
&9.328
&19.32
&37.42
&215.5
&1.003
 \\ \hline
7.0
&6.145
&9.329
&19.33
&37.52
&231.4
&1.003
 \\ \hline
8.0
&6.145
&9.330
&19.34
&37.58
&243.0
&1.003
 \\ \hline
9.0
&6.145
&9.330
&19.34
&37.62
&254.5
&1.003
 \\ \hline\hline
\end{tabular}\\

\vspace{10mm}

\pagebreak

Table IV. Ratio $(\Gamma/2)/(E/N)$ of $Na (F=1, m_F=1)$ atoms 
vs $\gamma$ and $B$ near the 
$B=907$G Feshbach resonance.
\vspace{10mm}

\begin{tabular}{llll}
\hline\hline
$\gamma$
&$B=906$G
&$B=906.5$G
&$B=908.5$G
 \\ \hline
0.1
&$2.1 \times 10^{-3}$
&$1.8 \times 10^{-2}$
&$1.2 \times 10^{-3}$
 \\ \hline
0.5
&$1.8 \times 10^{-3}$
&$1.2 \times 10^{-2}$
&$1.1 \times 10^{-3}$
 \\ \hline
9.0
&$1.8 \times 10^{-3}$
&$1.2 \times 10^{-2}$
&$1.1 \times 10^{-3}$
 \\ \hline\hline
\end{tabular}\\

\vspace{10mm}


TABLE V. Energy per particle of $Na (F=1, m_F=1)$ atoms, $E/N$, multiplied by
 $10^6 2 m a_{bg}^2/\hbar^2$, effective scattering length, $a_{eff}$ in units of $a_{bg}$, 
as a function of
 $\gamma$ at $B=907.1$G with $a(B)=-9 a_{bg}$.
\vspace{10mm}

\begin{tabular}{lll}
\hline\hline
$\gamma$
&$E/N$
&$a_{eff}$
 \\ \hline
0.1
&8.02
&2.59
 \\ \hline
0.5
&49.1
&13.8
 \\ \hline
1.0
&146
&31.8
 \\ \hline
1.5
&299
&49.5
 \\ \hline
2.0
&502
&72.6
 \\ \hline\hline
\end{tabular}\\
\vspace{10mm}

\pagebreak
\begin{figure}[ht]
\includegraphics{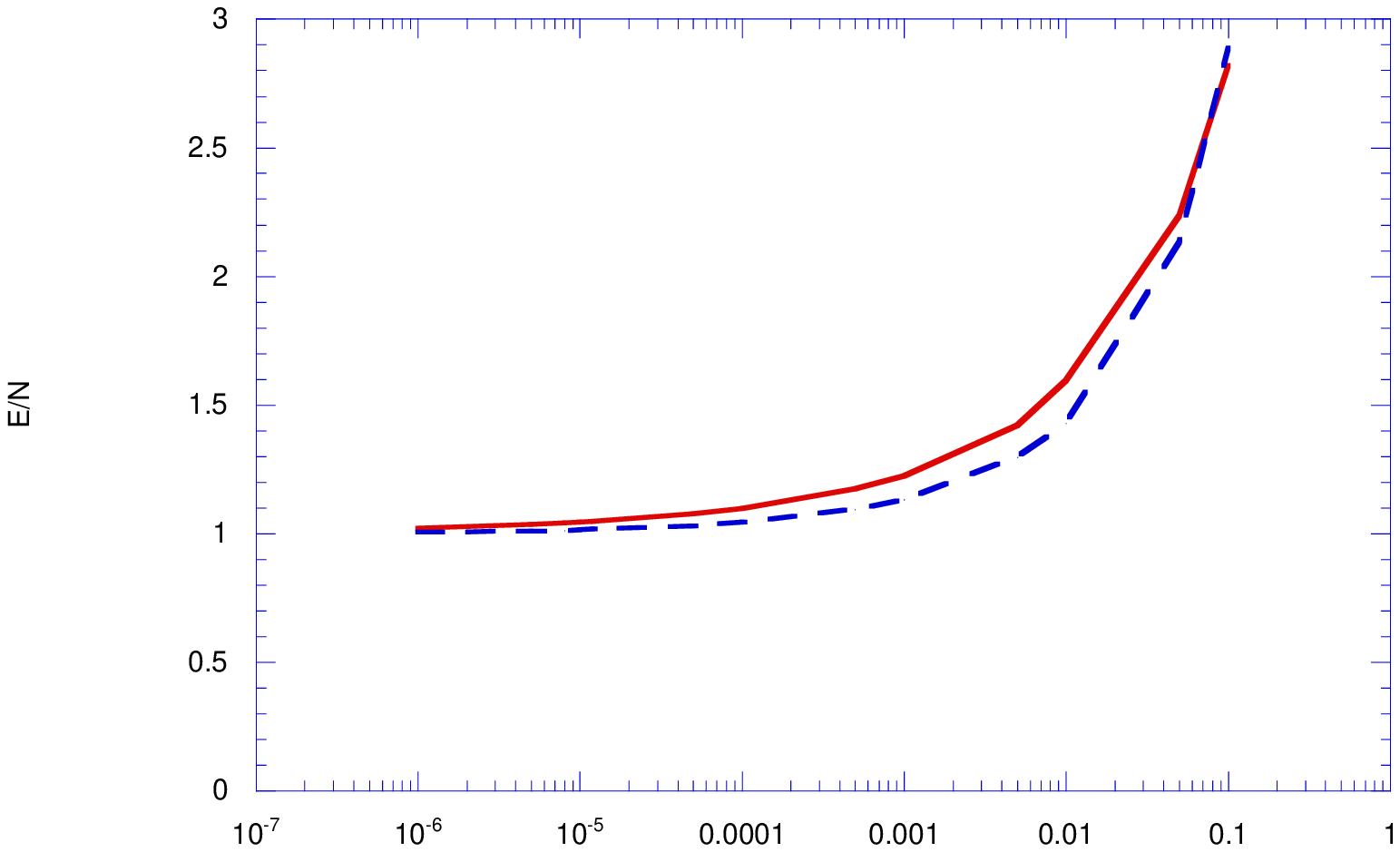}
\end{figure}
\begin{center}
$ a^3 n$
\end{center}

\vspace{10mm}
FIG.1. Ground-state energy per particles, $E/N$,in units of $2 \pi \hbar^2 n a/m$, vs $a^3 n$ for hard-sphere interactions. The LOCV results are shown by full line. The diffusion Monte-Carlo calculations [14] are shown as dashed line.
\pagebreak

\begin{figure}[ht]
\includegraphics{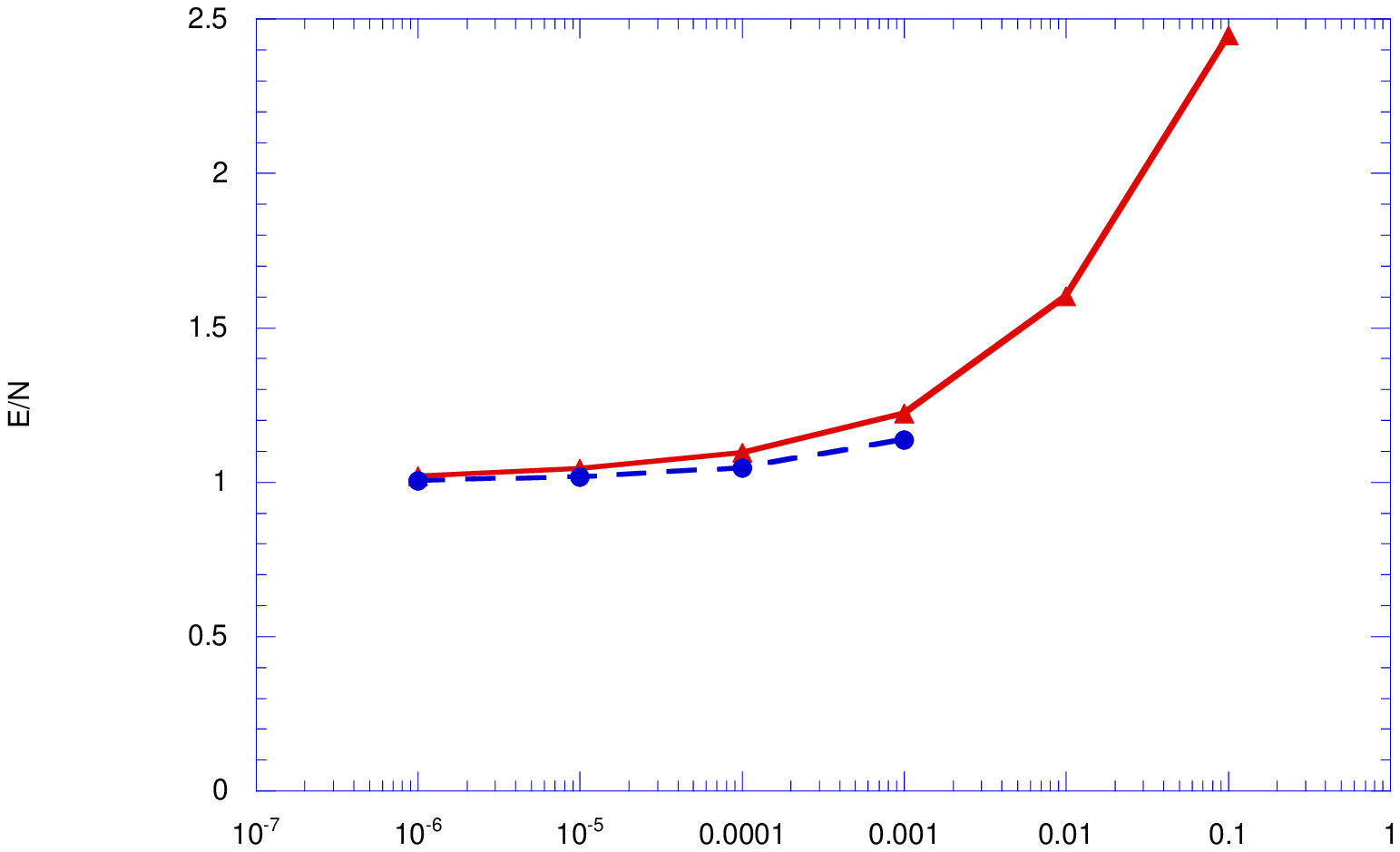}
\end{figure}
\begin{center}
$a^3 n$
\end{center}

\vspace{10mm}
FIG. 2. Ground-state energy per particles, $E/N$,in units of $2 \pi \hbar^2 n a/m
$, vs $a^3 n$ for hard-core square-well potential (HCSW).
Triangles and circles correspond to the LOCV and the diffusion Monte-Carlo [14]
results, respectively.
Lines are drawn  to  guide  the eyes.
\pagebreak

\begin{figure}[ht]
\includegraphics{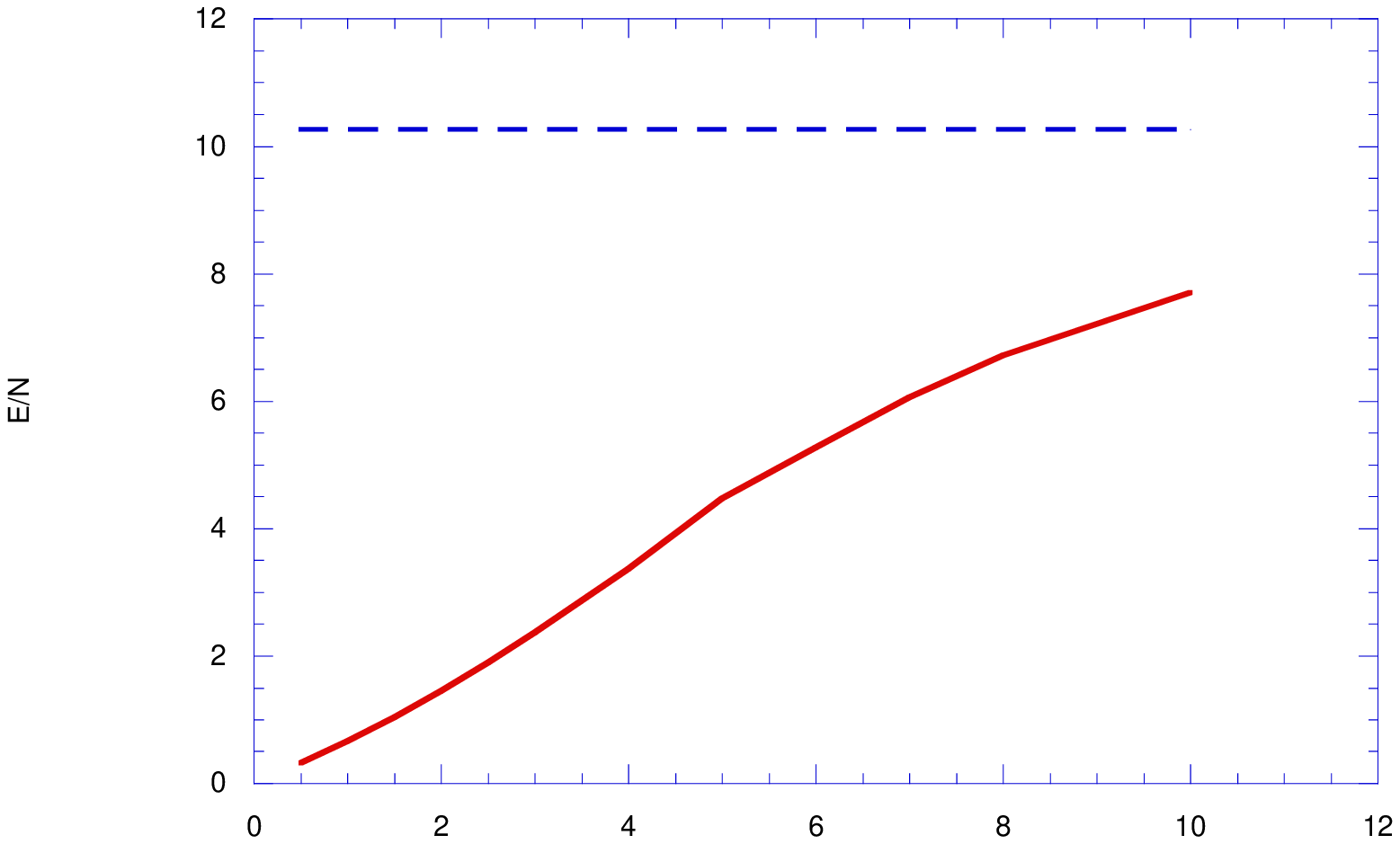}
\end{figure}
\begin{center}
${\bf \gamma}$
\end{center}

\vspace{10mm}
FIG. 3. Ground-state energy per particles, $E/N$, of $Na (F=1,m_F=+1)$ atoms
at 907 G in units of $10^{4} 2 m a_{bg}^2/\hbar^2$, as a function of parameter
$\gamma$, (solid line). Dashed line represent approximation of Ref.[6], Eq.(10).
($r_0=10^{2} a_{bg}$).

\pagebreak

{\bf REFERENCES}
\vspace{8pt}

\noindent
1. {\it Bose-Einstein Condensation in Atomic Gases}, Proceedings of the 
International
 School of Physics ``Enrico Fermi", edited by M. Inquscio, S. Stringari,
 and
 C.E. Wieman (IOS Press, Amsterdam, 1999);http://amo.phy.gasou.edu/bec.html;
http://jilawww.colorado.edu/bec/
 and references therein.

\noindent
2.  A.L. Fetter and A.A. Svidzinsky, J.Phys.: Condens. Matter { \bf13}, R135
(2001);
A.J. Leggett, Rev. Mod. Phys. {\bf 73}, 307 (2001);
 K. Burnett, M. Edwards, and C.W. Clark, Phys. Today, {\bf 52}, 37 (1999);
 F. Dalfovo, S. Giorgini, L. Pitaevskii, and S. Stringari, Rev. Mod. Phys.
 {\bf 71}, 463 (1999);
S. Parkins, and D.F. Walls, Phys. Rep. {\bf 303}, 1 (1998).

\noindent
3.  L.P. Pitaevskii, Zh. Eksp. Teor. Fiz. {\bf 40}, 646 (1961) [Sov. Phys. JETP
 {
\bf 13}, 451 (1961)];
   E.P. Gross, Nuovo Cimento {\bf 20}, 454 (1961); J. Math. Phys. {\bf 4}, 195 (
1963).

\noindent
4. S. Inouye, M.R. Andrews, J. Stenger, H.J. Miesner, D.M. Stamper-Kurn, and
 W.
Ketterle, Nature {\bf 392}, 151 (1998);
P. Courteille, R.S. Freeland, D.J. Heinzen, F.A. van Abeelen, and B.J. Verhaar,
Phys. Rev. Lett. {\bf 81}, 69 (1998);
J.L. Roberts, N.R. Claussen, J.P. Burke, C.H. Greene, E.A. Cornell, and
C.E. Wieman, Phys. Rev. Lett. {\bf 81}, 5109 (1998);
D.M. Stamper-Kurn, and  W. Ketterle, Phys. Rev. Lett. {\bf 82}, 2422 (1999);
J.L. Roberts, N.R. Claussen, S.L. Cornish, E.A. Donley, E.A. Cornell, and
C.E. Wieman, Phys. Rev. Lett. {\bf 86}, 4211 (2001);
E.A. Donley, N.R. Claussen, S.L. Cornish, J.L. Roberts, E.A. Cornell, and
C.E. Wieman, Nature {\bf 412}, 295 (2001).

\noindent
5. S.L. Cornish, N.R. Claussen, J.L. Roberts, E.A. Cornell, and  C.E. Wieman,
Phys. Rev. Lett. {\bf 85}, 1795 (2000).

\noindent
6.  S. Cowel, H. Heiselberg,
I.E. Morales, V.R. Pandharipande, and C.J. Pethick,
Phys. Rev. Lett. {\bf 88}, 210403 (2002).

\noindent
7. H. Heiselberg, Phys. Rev. A{\bf 63}, 043606 (2001).

\noindent
8. V.R. Pandharipande,
 Nucl. Phys. A{\bf 174}, 641 (1971);
V.R. Pandharipande, Nucl. Phys. A{\bf 178}, 123 (1971);
 V.R. Pandharipande and H.A. Bethe, Phys.
 Rev. C{\bf 7}, 1312 (1973);
V.R. Pandharipande and K.E.
Smith,
Phys. Rev. A{\bf 15}, 2486 (1977).

\noindent
9. A.I. Baz', Ya.B. Zel'dovich, and A.M. Perelomov, {\it Scattering, Reaction, and Decays in Nonrelativistic Quantum Mechanics} [in Russian] (Nauka, Moscow 1971); 
Yu. N. Demkov and V.N. Ostrovskii, {\it Zero-Range Potentials and their Applications in Atomic Physics} (Plenum Press, New York 1988).

\noindent
10. W. Lenz, Z. Phys. {\bf 56}, 778 (1929).

\noindent
11. T.D. Lee, K.Huang, and C.N. Yang, Phys. Rev. {\bf 106},
 1135 (1957).

\noindent
12.  T.T. Wu, Phys. Rev. {\bf
115}, 1390 (1959).

\noindent
13.  E.
Braaten, H.-W. Hammer, and T. Mehen, Phys. Rev. Lett. {\bf 88}, 040401 (2002)
and references therein.

\noindent
14.
 S. Giorgini, J. Boronat, and J. Casulleras, Phys. Rev. A{\bf 60}, 5129 (1999).

\noindent
15. Y.E. Kim and A.L. Zubarev, cond-mat/0205422.

\noindent
16.  G.S. Nunes, J. Phys. B: At. Mol. Opt. Phys. {\bf 32}, 4293 (1999)
 and
references therein.

\noindent
17.  A. Fabrocini and A. Polls,
Phys. Rev. A{\bf 60}, 2319 (1999).

\noindent
18. A. Fabrochini and A. Polls, Phys. Rev. A{\bf 64}, 063610 (2001).

\noindent
19. L. Pitaevskii and S. Stringari, Phys. Rev. Lett.
{\bf 81}, 4541 (1998).

 \noindent
20. G. Baym, J. Phys. B: At. Mol. Opt. Phys. {\bf 34}, 4541 (2001).

\noindent
21. E. Braaten, H.-W. Hammer, and S. Hermans, Phys. Rev. A{\bf 63}, 063609 (2001).

\noindent
22. K. Huang, {\it Statistical Mechanics} (Wiley, New York, 1987).

\noindent
23. J.H. Macek, Few-Body Systems {\bf 31}, 241 (2002);
O.I. Kartavtsev and J.H. Macek, Few-Body Systems {\bf 31}, 249 (2002).

\noindent
24. A.J. Moerdijk, B.J. Verhaar, and A. Axelsson, Phys. Rev. A{\bf 51}, 4852 (1995).

\noindent
25. J. Stenger, S. Inouye, M.R. Andrews, H.-J. Meisner, D.M. Stamper-Kurn, and W. Ketterle, Phys. Rev. Lett. {\bf 82}, 2422 (1999).

\noindent
26. S.J.J.M.F. Kokkelmans, J.N. Milstein, M.L. Chiofallo, R. Walser, and M.J. 
Holland, Phys. Rev. A{\bf 65}, 053617 (2002).

\noindent
27. E. Timmermans, P. Tommasini, M,  Hussein, and A. Kerman,
Phys. Rep. {\bf 315}, 199 (1999).

\noindent
28. V.A, Yurovsky, A. Ben-Reuven, P.S.  Julienne,  and C.J. Williams,
Phys. Rev. A{\bf 60}, R765 (1999).

\noindent
29. F.A. van Abeelen  and B.J.  Verhaar, Phys. Rev. Lett. {\bf 83},1550 (1999).

\noindent
30. D. Blume and C.H.  Greene, Phys. Rev. A{\bf 66}, 013601 (2002).

\noindent
31.  D.M. Gangardt and  G.V. Shlyapnikov, cond-mat/0207338.
\end{document}